\documentclass[aps,prev,twocolumn,preprintnumbers,floatfix,nofootinbib]{revtex4-1}
\pdfoutput=1

\newcommand{\be}{\begin{equation}}
\newcommand{\ee}{\end{equation}}
\newcommand{\bea}{\begin{eqnarray}}
\newcommand{\eea}{\end{eqnarray}}

\usepackage{mathrsfs}
\usepackage{amsmath, amsthm, amssymb}
\usepackage{color}
\usepackage{epstopdf}
\usepackage[pdftex]{graphicx}
\usepackage{amssymb}
\usepackage{verbatim}
\usepackage{hyperref}
\usepackage{enumerate}

\begin{document}

\preprint{KCL-PH-TH/2017-03}

\title{Searching for the QCD Axion with Gravitational Microlensing}

\author{Malcolm Fairbairn$^{a}$}
\email{malcolm.fairbairn@kcl.ac.uk}
\author{David J. E. Marsh$^{a}$}
\email{david.marsh@kcl.ac.uk}
\author{J\'er\'emie Quevillon$^{a}$}
\email{jeremie.quevillon@kcl.ac.uk}

\vspace{1cm}
\affiliation{
${}^a$ KingÕs College London, Strand, London, WC2R 2LS, United Kingdom}

\begin{abstract}

The phase transition responsible for axion dark matter production can create large amplitude isocurvature perturbations which collapse into dense objects known as axion miniclusters. We use microlensing data from the EROS survey, and from recent observations with the Subaru Hyper Suprime Cam to place constraints on the minicluster scenario. We compute the microlensing event rate for miniclusters treating them as spatially extended objects.  Using the published bounds on the number of microlensing events we bound the fraction of DM collapsed into miniclusters, $f_{\rm MC}$. For an axion with temperature dependent mass consistent with the QCD axion we find $f_{\rm MC}<0.083(m_a/100\,\mu\text{eV})^{0.12}$, which represents the first observational constraint on the minicluster fraction. We forecast that a high-efficiency observation of around ten nights with Subaru would be sufficient to constrain $f_{\rm MC}\lesssim 0.004$ over the entire QCD axion mass range. We make various approximations to derive these constraints and dedicated analyses by the observing teams of EROS and Subaru are necessary to confirm our results. If accurate theoretical predictions for $f_{\rm MC}$ can be made in future then microlensing can be used to exclude, or discover, the QCD axion. Further details of our computations are presented in a companion paper~\cite{FMQR}.

\end{abstract}

\maketitle

The QCD axion~\cite{weinberg1978,wilczek1978,1979PhRvL..43..103K,1980NuPhB.166..493S,Zhitnitsky:1980tq,1981PhLB..104..199D} remains one of the most well-motivated and viable candidates for particle dark matter (DM).  The axion is a pseudo-Nambu-Goldstone boson of a spontaneously broken global $U(1)$ symmetry, known as a Peccei-Quinn (PQ) symmetry~\cite{pecceiquinn1977}. PQ symmetry breaking occurs when the temperature of the Universe drops below the symmetry breaking scale $T_{\rm PQ}\sim f_a$. The cosmology of the axion is determined by the cosmic epoch during which symmetry breaking occurs~\cite{2008LNP...741...19S,Marsh:2015xka}.  If the PQ symmetry is broken after smooth cosmic initial conditions are established (by, for example, inflation) then topological defects and large amplitude axion field fluctuations will be present on scales of order the horizon size at symmetry breaking~\cite{Hogan:1988mp,Sikivie:1982qv,1987PhLB..195..361H}. For models of inflation, the observational bound on the cosmic microwave background tensor to scalar ratio of $r_T\leq 0.07$~\cite{2016PhRvL.116c1302B} implies this scenario for symmetry breaking is only possible for $f_a\lesssim 10^{13}\text{ GeV}$. 
\begin{figure}
\vspace{-0.2em}\includegraphics[width=0.9\columnwidth]{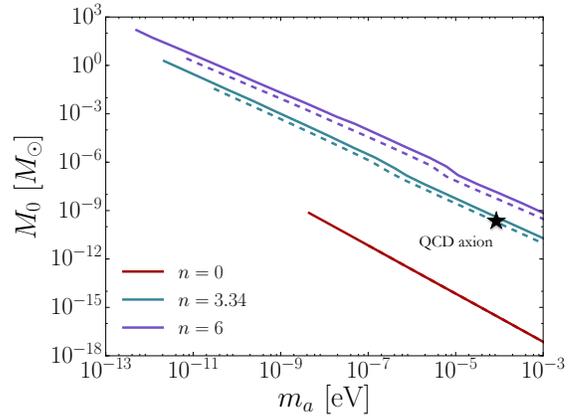}
\caption{{\bf The Characteristic Minicluster Mass}: We plot $M_0$, as a function of the axion mass, $m_a$, for different temperature evolutions of the axion mass parameterised by index $n$. Solid lines show the most realistic assumptions about the relic density, while dashed lines relax those assumptions slightly. When the axion mass is temperature independent ($n=0$), the two scenarios are equivalent for minicluster mass. Lines terminate at a lower bound on $m_a$ set by the DM relic abundance and the constraint $f_a\lesssim 10^{13}\text{ GeV}$ for minicluster production.}  
\label{fig:M0_mass}
\end{figure}

The Kibble mechanism~\cite{1976JPhA....9.1387K} smoothes the axion field on the horizon scale until such a time that the axion mass becomes cosmically relevant: $3H(T_0)\approx m_a(T_0)$, where $H(T)$ is the Hubble rate and we have allowed temperature dependence of the axion mass. At this epoch, the topological defects decay (we consider only the case with domain wall number equal to unity)~\cite{2015PhRvD..91f5014K}, and the axion field is left with large amplitude isocurvature fluctuations on the horizon scale.

Once cosmological structure begins to grow at matter-radiation equality, the isocurvature perturbations are converted into curvature perturbations, and promptly collapse into dense bound structures of DM known as \emph{axion miniclusters}~\cite{Hogan:1988mp,1993PhRvL..71.3051K,1994PhRvD..49.5040K,Kolb:1994fi,Kolb:1995bu,2007PhRvD..75d3511Z,2016arXiv160900208H}. The characteristic minicluster mass, $M_0$, is given by the total mass of DM contained within the horizon at the epoch $T_0$:
\be
M_0 = \bar{\rho}_a\frac{4}{3}\pi \left(\frac{\pi}{a(T_0)H(T_0)}\right)^3\, ,
\ee
where $a$ is the cosmic scale factor of the Friedmann-Lema\^itre-Robertson-Walker metric, and we have considered a spherical patch of radius $R=\pi/k_0$ for comoving wavevector $k_0=a(T_0)H(T_0)$ (here and throughout $\hbar=c=1$). The definition of $M_0$ depends upon filtering of the mass function~\cite{FMQR}. Ours differs from others in the literature that take a cubic volume $\sim k_0^{-3}$.

The temperature $T_0$ sets the time when the axion field goes from having equation of state $w=-1$ to $w=0$, and therefore depends on the temperature evolution of the axion mass, $m_a(T)=m_{a,0}(T/T_c)^{-n}$, with $m_a(T<T_c)=m_{a,0}\equiv m_a$. The index $n$ parameterizes the sharpness of the phase transition, and the critical temperature $T_c\approx \sqrt{m_af_a}$ (for the QCD axion $T_c\approx \Lambda_{\rm QCD}\approx 200\text{ MeV}\approx 2.5\sqrt{m_af_a}$; the case $T_c\gg \sqrt{m_af_a}$ occurs for some axion-like particles~\cite{2014JHEP...06..037D} and is equivalent to $n=0$). This phase transition also determines the axion DM density~\cite{1983PhLB..120..127P,1983PhLB..120..133A,1983PhLB..120..137D}. Fixing the DM density $\Omega_ch^2=0.12$~\cite{planck_2015_params} determines an $n$-dependent relationship between $m_a$ and $f_a$, such that $M_0=M_0(m_a,n)$. 

Following the standard computation for the axion DM density~\cite{Wantz:2009it}, and accounting for uncertainties due to anharmonicities in the axion potential and the decay of topological defects~\cite{2015PhRvD..91f5014K}, we compute $M_0(m_a,n)$ for various $n$ (see Fig.~\ref{fig:M0_mass}). As a representative of the QCD axion we take $n\approx 3.34$ from the ``interacting instanton liquid'' model for the QCD topological susceptibility~\cite{Wantz:2009it}, which is consistent with the results from lattice simulations ($n\approx 3.55\pm 0.30$~\cite{2016PhLB..752..175B,Borsanyi:2016ksw}). 
\begin{figure}
\center
\includegraphics[width=0.9\columnwidth]{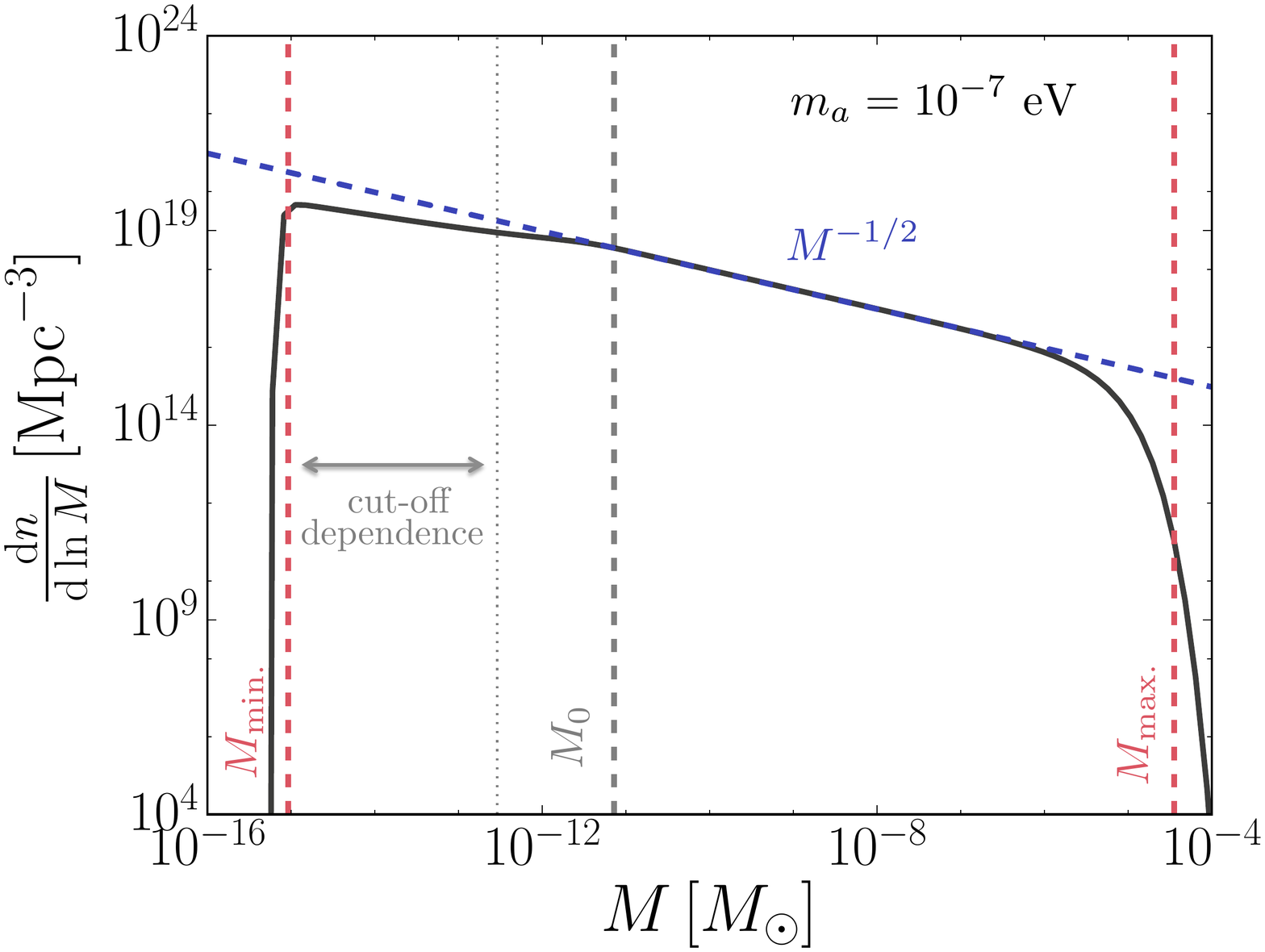}
\caption{{\bf Parametrization of the Minicluster Mass Function}. The mass function can be well fit by two cut-offs and a single slope parameter, $M^{-1/2}$, derived from white noise initial conditions cut at $M_0$. For the numerical calculation (solid line), the normalization is fixed to be per unit volume. For the substructure mass function, we normalize by $f_{\rm MC}$. For illustration we take $m_a=10^{-7}\text{ eV}$ and $n=0$ for the axion mass temperature dependence.}
  \label{parametrize_HMF}
\end{figure}

After their initial formation, miniclusters of mass $M_0$ undergo hierarchical structure formation, and \emph{collapse into larger ``minicluster halos'' (MCHs)} as substructure within larger galactic halos. Miniclusters collapse much earlier and on different scales than galactic halos, and so we treat these two periods of structure formation independently.

Hierarchical structure formation can be computed following the Press-Schechter~\cite{Press:1973iz} approach as shown in Fig.~\ref{parametrize_HMF}. The slope of the MCH mass function is fixed by the (cut white noise) initial conditions giving a mass variance $\sigma^2(M)\propto M^{-1}$ for $M\gtrsim M_0$, constant $M\lesssim M_0$. The maximum MCH mass is determined by linear growth and the Gaussian cut-off for crossing the collapse threshold, dropping to one percent at $M\approx 5\times 10^6 M_0$. The minimum axion halo mass is seen in simulations~\cite{2007PhRvD..75d3511Z,Schive:2014dra,2016arXiv161105892C} and is determined by a combination of the initial conditions (Kibble mechanism) and the axion Jeans scale/de Broglie wavelength~\cite{khlopov_scalar,Marsh:2013ywa} and is cut-off dependent. MCHs with $M\ll M_0$, however, play little role in microlensing for the QCD axion for the surveys considered.

We normalize the substructure mass function to
\be
f_{\rm MC}= \frac{1}{M_{\rm host}}\int M\frac{{\rm d}n}{{\rm d}M}{\rm d}M \, ,
\ee
for host galaxy mass $M_{\rm host}$ and minicluster fraction $f_{\rm MC}$. The presence of $f_{\rm MC}$ as a free parameter accounts for the fact that, due to the axion population from topological defect decay and the effects of e.g. tidal stripping~\cite{2016JCAP...01..035T}, only a fraction of axions end up bound in miniclusters.

In some cases miniclusters and MCHs can be massive enough and dense enough to impact gravitational microlensing. Thus, searches for axion miniclusters are related to searches for non-particle DM candidates such as MAssive Compact Halo Objects (MACHOs)~\cite{1986ApJ...304....1P,Griest:1990vu}, and primordial black holes (PBHs, e.g. Refs~\cite{Green:2016xgy,2017arXiv170102151N}).

We compute the lensing signal for miniclusters treating them as extended objects. Miniclusters can remain \emph{isolated} from each other as they join larger haloes, form \emph{dense MCHs}, or become disrupted into \emph{diffuse MCHs}. We consider all of these possibilities below and in more detail in Ref.~\cite{FMQR}. The true model of structure formation with miniclusters must be determined by simulations: our models bracket the possibilities.

We computed the gravitational microlensing signal from axion miniclusters, and MCHs, for the EROS survey of the Large Magellanic Cloud (LMC)~\cite{Tisserand:2006zx} and for the Subaru Hyper Suprime Cam (HSC) survey of Andromeda (M31)~\cite{2017arXiv170102151N}. EROS has a high microlensing efficiency for timescales between one day and 1000 days, while HSC observations have high efficiency for timescales between two minutes and seven hours. Thus the two surveys probe different characteristic lens masses~\cite{Griest:1990vu}. We make various approximations in order to handle the constraints from these surveys in a simple manner, and emphasize that a dedicated analysis by observers is desirable. 

\emph{Microlensing with Miniclusters:} A key quantity in gravitational microlensing is the Einstein radius:
\be
R_{\rm E}(x,M) = 2\left[GMx(1-x)d_s\right]^{1/2}\, ,
\ee
where $M$ is the lens mass, $d_s$ is the distance from the observer to the source, and $x=d/d_s$ where $d$ is the distance from the observer to the lens. For a point-like lens, the Einstein radius defines the shape of the ``microlensing tube''~\cite{Griest:1990vu}. This is the volume within which a lens must pass for the lensing amplification, $A$, to exceed 1.34, $A=1.34$ being the threshold applied to the lightcurves in Refs.~\cite{Tisserand:2006zx,2017arXiv170102151N}.

Miniclusters are extended objects with scale radius determined by the characteristic density. The characteristic density of a minicluster found in numerical simulations is~\cite{1993PhRvL..71.3051K,1994PhRvD..49.5040K,Kolb:1994fi,Kolb:1995bu}:
\be
\rho_c=140\delta^3(1+\delta)\rho_a (1+z_{\rm eq})^3 \, ,
\label{eqn:density_char}
\ee
where $\rho_a$ is the cosmic axion DM density, and $z_{\rm eq}$ is the redshift of matter radiation equality (these and the other cosmological parameters we use are determined by the cosmic microwave background anisotropies~\cite{planck_2015_params}). The parameter $\delta$ is the characteristic overdensity of a minicluster at the time of formation. In numerical simulations miniclusters are observed to have a distribution of values for $\delta$ given by ${\rm d}n/{\rm d}\delta$ which we take from the numerical results of Fig.~2 in Ref.~\cite{Kolb:1995bu}, and which we fit with a Pearson-VII distribution to extend to large $\delta$.

Above the axion de Broglie wavelength (which can be safely neglected in the density profiles for microlensing~\cite{FMQR}), we treat the density profiles of miniclusters/MCH as Navarro-Frenk-White (NFW)~\cite{Navarroetal1997} type.  Eq.~\eqref{eqn:density_char} defines the NFW characteristic density $\rho(r)=\rho_c/[(r/r_s)(1+r/r_s)^2]$, with the scale radius defined from $\rho_c$ after fixing the total mass, $M$, of the minicluster/MCH and assuming the profiles extends to 100 $r_s$. An alternative minicluster/MCH density profile fixes $\rho_c$ as the core density and the radial dependence as $r^{-9/4}$ from self-similar infall~\cite{2017arXiv170103118O} and is explored in Ref.~\cite{FMQR}.

We integrate the three dimensional density profile along the line of sight towards the centre of the halo to obtain a surface density for lensing. We then calculate the magnification for an axisymmetric mass distribution with impact parameter $\ell$ from the line of sight
\begin{align}
&A=\left[\left(1-B\right)\left(1+B-C\right)\right]^{-1} \, , \\
C=\frac{1}{\Sigma_c \pi \ell}&\frac{{\rm d}M(\ell)}{{\rm d}\ell}\,\,; \,\, B=\frac{M(\ell)}{\Sigma_c \pi \ell^2}\,\, ; \,\, \Sigma_c=\frac{1}{4\pi G d_s x(1-x)} \nonumber \, .
\end{align}
In this way we compute the shape of the microlensing tube given by the value of $\ell$ corresponding to a magnification of $A=1.34$ for a minicluster defined by $(M,\delta)$. 

From our numerical lensing calculations, we find that the shape of the microlensing tube is still reasonably well described by $R_{\rm E}(x,M)$, but with a rescaling factor, $\mathcal{R}$, that depends on $\delta$ and $M$~\cite{FMQR}, such that the minicluster microlensing tube is given by: 
\be
R_{\rm MC}(x,M,\delta)=\mathcal{R}(\delta,M)R_E(x,M) \, .
\ee
When a  mincluster/MCH is diffuse, the tube is smaller. There is a minimum value of $\delta$ below which there is no value of impact parameter $\ell$ for which $A\geq 1.34$, i.e. $\mathcal{R}(\delta<\delta_{\rm min})=0$ with $\delta_{\rm min.}=\delta_{\rm min.}(M)$ given approximately by $r_s/R_E>1$. This reduces considerably the expected number of microlensing events for miniclusters compared to point masses (MACHOs, PBHs). For $\delta\gg \delta_{\rm min}$ the limiting behaviour is that of a point mass, $\mathcal{R}\rightarrow 1$.

The rate of microlensing events of duration $\hat{t}$ for miniclusters is:
\be
\hspace{-0.1in}\frac{{\rm d}\Gamma}{{\rm d}\hat{t}}= \frac{32 d_s}{\hat{t}^4v_c^2}\int_0^\infty \left\{\frac{{\rm d}n}{{\rm d}\delta}\int_0^\infty\left[\frac{{\rm d}n}{{\rm d}M}\int_0^1\rho_{\rm DM}R_{\rm MC}^4e^{-Q}{\rm d }x\right]{\rm d}M\right\}{\rm d}\delta \, ,
\ee
where $v_c\approx 220\text{ km s}^{-1}$ is the local circular speed, $\rho_{\rm DM}$ is the line of sight DM density to the source and we have suppressed the dependencies on, $x$, $M$ and $\delta$ of the integrand. The factor $e^{-Q}$ with $Q=4R_{\rm MC}^2/(\hat{t}^2v_c^2)$ emerges by approximating the Bessel function in the lensing integral~\cite{Griest:1990vu,1996ApJ...461...84A}.

The EROS survey observed the LMC, at a distance $d_{\rm LMC}=50\text{ kpc}$, considering only lensing events of LMC stars by DM in the Milky Way (MW). EROS models the MW as a cored isothermal sphere:
\be
\rho_{\rm MW,EROS}(r)=\rho_0\frac{R_c^2+R_\oplus^2}{R_c^2+r^2}\, ,
\ee
where $R_\oplus=8.5\text{ kpc}$ is radial distance of the Earth in the MW, $R_c=5\text{ kpc}$ and $\rho_0=0.0079\, M_\odot\text{pc}^{-3}$. A minicluster at distance $d$ from Earth on the line of sight to the LMC has radial coordinate in the MW $r_{\rm MW}^2(d)=R_\oplus^2-2R_\oplus d\cos l_{\rm LMC}\cos b_{\rm LMC}+d^2$, where $(l,b)$ are the measured Galactic coordinates.

HSC observed Andromeda (M31). For such an observation, one must consider lensing of stars in M31 by both DM in the MW and in M31 itself, and the event rate is given by ${\rm d}\Gamma={\rm d}\Gamma_{\rm MW}+{\rm d}\Gamma_{\rm M31}$. HSC model both the MW and M31 as NFW profiles with halo parameters from Ref.~\cite{2002ApJ...573..597K} quoted in Ref.~\cite{2017arXiv170102151N}. M31 is at a distance $d_{\rm M31}=770\text{ kpc}$. A minicluster in the MW at distance $d$ from Earth on the line of sight to M31 has radial coordinate $r_{\rm MW}(d)^2=R_\oplus^2-2R_\oplus d \cos l_{\rm M31}\cos b_{\rm M31}+d^2$, while a minicluster in M31 at a distance $d$ from Earth has radial coordinate in M31 $r_{\rm M31}(d)\approx d_{\rm M31}-d$. Note we have not taken into account the reduction in sensitivity below about $10^{-9} M_\odot$ due to the Einstein radius subtending an angle less than that of the diameter of the source star.  This could reduce our sensitivity in the most interesting region by a factor of a few~\cite{2017arXiv170102151N}.

The number of expected microlensing events is:
\be
N_{\rm exp}=E\int {\rm d}\hat{t}\frac{{\rm d}\Gamma}{{\rm d}\hat{t}} \epsilon (\hat{t})\, ,
\ee
where $E$ is the total exposure in star-years and $\epsilon (\hat{t})$ is the microlensing efficiency of the survey. EROS give the microlensing efficiency in Fig.~11 of Ref.~\cite{Tisserand:2006zx}, which we digitize. The exposure is as $E_{\rm EROS}=3.68\times 10^7$ star-years~\cite{Tisserand:2006zx}. 

HSC use a Monte Carlo technique to determine the efficiency in each region of the observing field, and separately for different magnitudes of stars. Reproducing such an analysis is beyond the scope of our work and so we make a series of approximations to obtain HSC constraints. We model the efficiency from Fig.~14 of Ref.~\cite{2017arXiv170102151N} as a step function with $\epsilon=0.5$ between the sampling of two minutes and the observing time of seven hours. We normalise the exposure to reproduce the bound on the PBH fraction (e.g. Fig. 21 of Ref.~\cite{2017arXiv170102151N}) using our methods.

\begin{figure}
\includegraphics[width=\columnwidth]{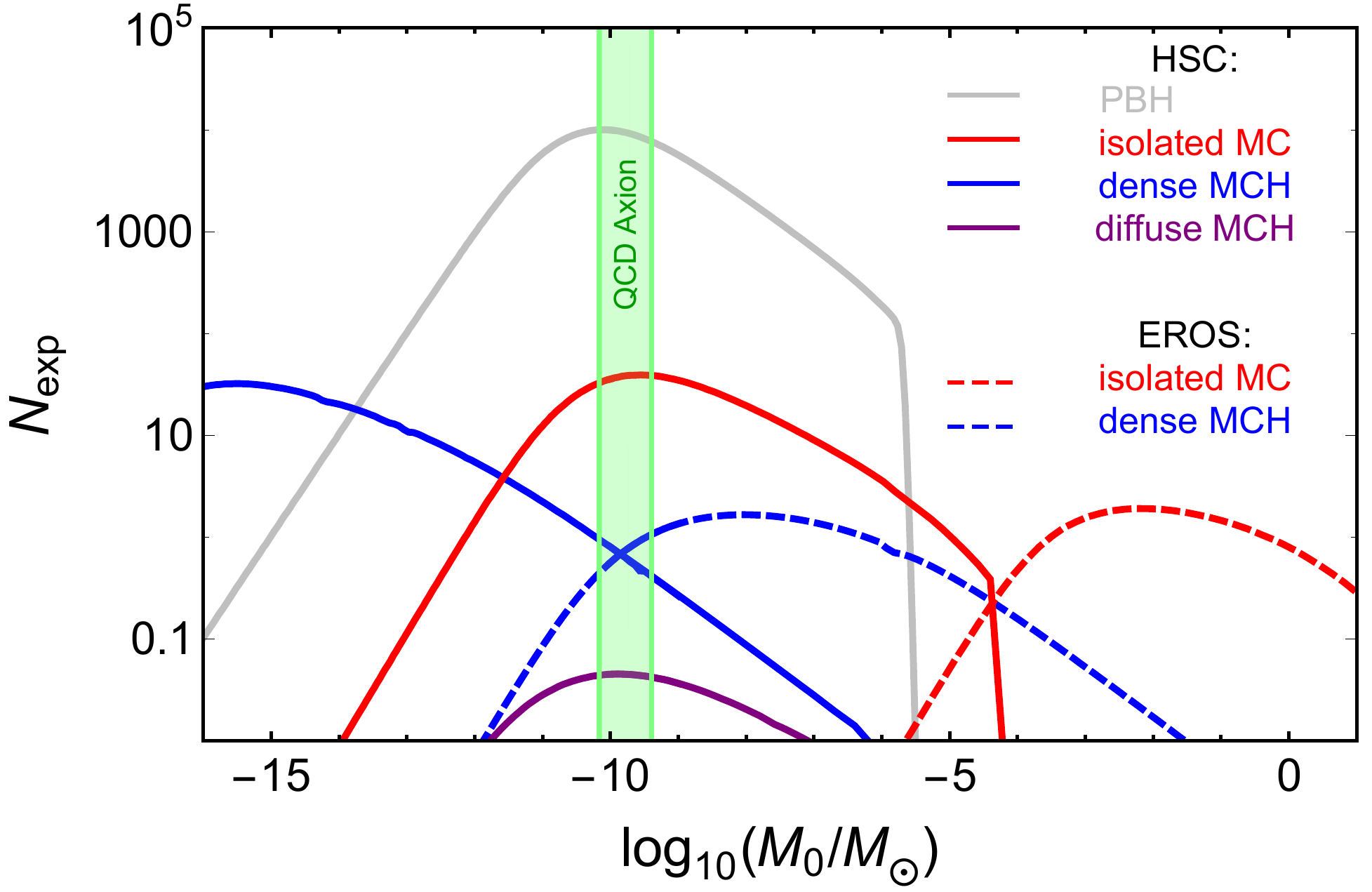}
\vspace{-2.5em}\caption{{\bf Expected Microlensing Events}: Here we assume that all the DM is composed of miniclusters on small scales. Lines show the effects of our modelling of the minicluster mass function and density profile for HSC and the EROS survey.}  
\label{fig:Nexp_M0}
\end{figure}

\emph{Results:} We show the expected number of microlensing events in the minicluster scenario as a function of $M_0$ in Fig.~\ref{fig:Nexp_M0} for HSC and EROS with $f_{\rm MC}=1$. The number of events in HSC is far larger than for EROS due to the huge volume of DM between Earth and M31 leading to a larger optical depth to microlensing for HSC~\cite{2017arXiv170102151N}. To show the effects of our modelling we show four different calculations of $N_{\rm exp.}$ for HSC. 

In the first, we compute the event rate for PBHs (i.e point like object) of fixed mass $M_0$ (i.e. Dirac-delta-function mass distribution) to normalise the exposure and efficiency. 

We then compute the case of isolated miniclusters, with density profiles determined by ${\rm d}n/{\rm d}\delta$. This reduces the number of events by a factor of $\mathcal{O}(10^2)$ due to the requirement of large $\delta$ such that $\mathcal{R}>0$. We consider this most conservative: miniclusters are too dense to suffer much disruption on mergers, and MCHs are likely to be a ``plum pudding'' of $M_0$ objects. In this case, for the HSC cadence and QCD axion, the modulating role of the MCH mass function is not relevant
 
The dense MCH case includes in addition the effects of ${\rm d}n/{\rm d}M$. A microlensing survey is sensitive to objects of fixed mass $M$. The mass function spreads the MCHs to $M>M_0$ (with more total mass at larger $M$), shifting the central $M_0$ to smaller values. The density profiles of the dense MCHs are also computed using ${\rm d}n/{\rm d}\delta$ i.e. mergers forming MCHs are assumed to preserve the distribution of halo concentrations.

Finally, the diffuse minicluster case uses ${\rm d}n/{\rm d}M$, but assumes that all MCHs with $M$ outside a small window near $M_0$ have too low density for microlensing. The cut in ${\rm d}n/{\rm d}M$ reduces the number of events. This is the most pessimistic model, corresponding to an effective reduction in $f_{\rm MC}$ caused by mergers.

\begin{figure}
\includegraphics[width=\columnwidth]{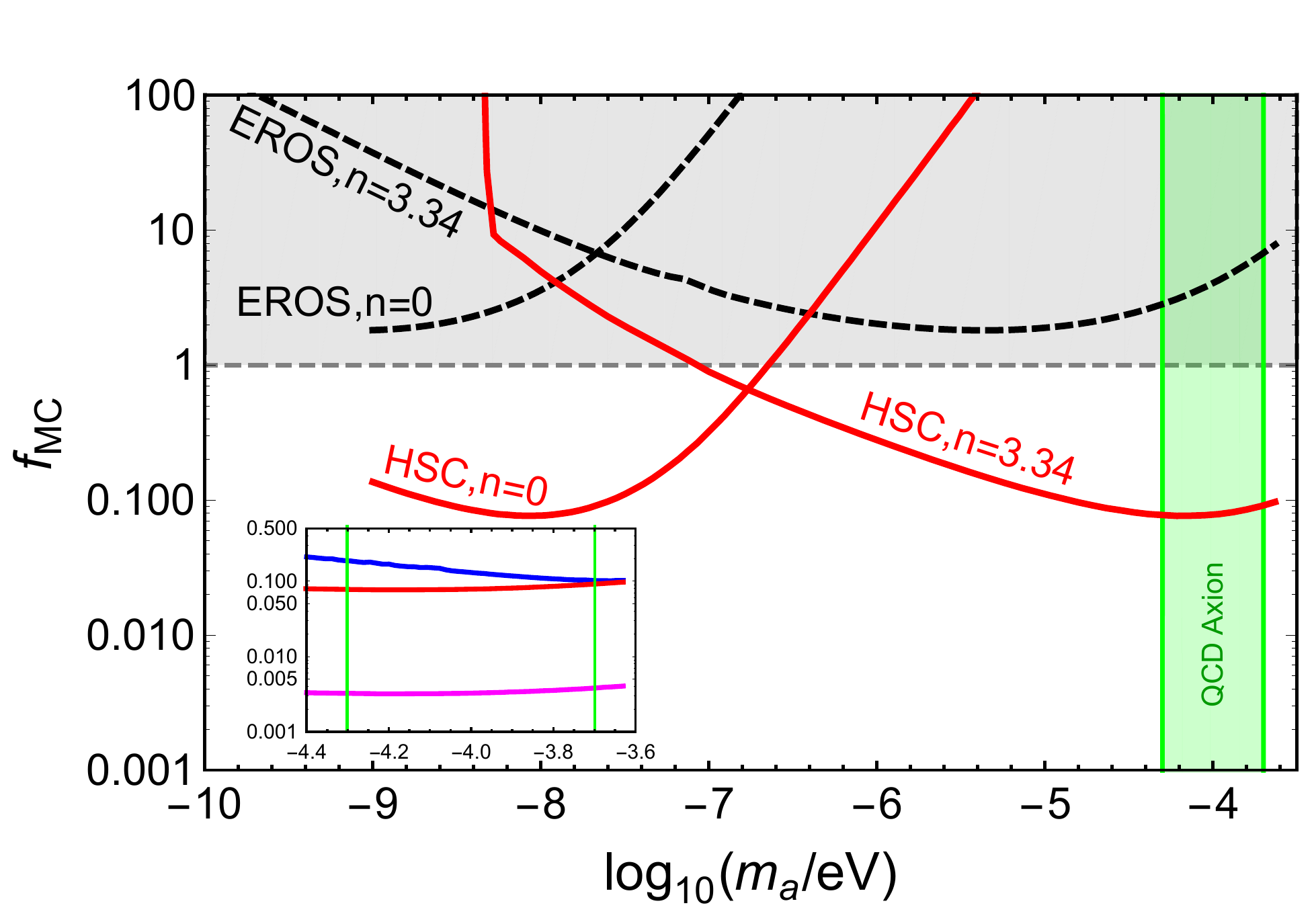}
\vspace{-2.5em}\caption{{\bf Limits on the Fraction of DM collapsed into Miniclusters}: The model adopted is for ``isolated miniclusters'', which we consider the most realistic. The shaded region shows the allowed mass for the QCD axion in this scenario. Where the $n=3.34$ lines intersect this region, $f_{\rm MC}$ is constrained for the QCD axion. The magenta (blue) line in the inset shows a hypothetical improved observation by HSC ten nights with an efficiency $\epsilon\sim 1$ in the case of isolated miniclusters (dense MHCs).}  
\label{fig:fMC_ma}
\end{figure}

Taking both EROS and HSC to have observed zero microlensing candidates the Poisson statistics 95\% C.L. limit on the number of expected events is $N_{\rm exp}\leq 3$~\cite{Tisserand:2006zx,2017arXiv170102151N}. Using this limit we find the constraints on $f_{\rm MC}$ as a function of axion mass $m_a$ presented in Fig.~\ref{fig:fMC_ma}. We find that EROS is unable to place a bound on $f_{\rm MC}<1$. 

HSC, on the other hand, does. The shaded band shows the allowed mass for the QCD axion fixed by $m_a=6.6\,\mu\text{eV}(10^{12}\text{ GeV}/f_a)$~\cite{weinberg1978,wilczek1978} and the relic density: $50\,\mu\text{eV}\lesssim m_a\lesssim 200\,\mu\text{eV}$~\cite{2016arXiv161001639B}. The solid lines show the HSC constraint: where the $n=3.34$ line intersects the shaded band, $f_{\rm MC}$ is bounded for the QCD axion, and we find $f_{\rm MC}<0.083(m_a/100\,\mu\text{eV})^{0.12}$ in the isolated miniclusters case.

These results could be improved as shown in Fig.~\ref{fig:fMC_ma} (inset) where the magenta line shows a hypothetical improved observation by HSC, extending to ten nights with an efficiency $\epsilon\sim 1$, leading to a forecast bound of $f_{\rm MC} \lesssim 0.004$ for the QCD axion in the isolated miniclusters case. The improved observation would also able to bound $f_{\rm MC} \lesssim 0.1$ in the more pessimistic dense MCH scenario. We advocate a dedicated analysis of the HSC microlensing data to place more rigorous bounds on $f_{\rm MC}$ than we have approximated, and for a longer microlensing survey in order to improve those bounds further. Ref.~\cite{FMQR}, includes the necessary light curves. Ref.~\cite{FMQR} also discusses various theoretical uncertainties and modelling that can give small shifts in the constraints. The largest uncertainty comes from our simplified modelling of the lensing efficiency. We are confident, however, that a more thorough analysis by the observing teams will show that HSC, and microlensing in general, is now a powerful tool to constrain the QCD axion.

In this paper we have used microlensing to place the first observational bounds on the DM axion minicluster fraction, $f_{\rm MC}$. This quantity is poorly understood theoretically, and naively could be of order unity. If the minicluster fraction were unity then axion DM detection in the lab~\cite{2015ARNPS..65..485G} in this mass range , e.g. by ``MADMAX''~\cite{2016arXiv161105865T}, would be much more difficult due to the small probability of an encounter between the Earth and a minicluster. Constraining $f_{\rm MC}$ observationally is an important task. 

If axions are ever detected directly in the lab then tidal stripping of miniclusters allows $f_{\rm MC}$ to be measured from the phase-space distribution~\cite{2016JCAP...01..035T,2017arXiv170103118O}. Independently of $f_{\rm MC}$, axions in the mass range accessible to microlensing can be detected via the force they mediate using the proposed experiment ``ARIADNE''~\cite{2014PhRvL.113p1801A}.

If accurate theoretical predictions for $f_{\rm MC}$ are made through numerical simulation then our results and future microlensing surveys could be used to exclude the existence of the QCD axion, or indeed discover evidence for it.

\paragraph*{Acknowledgments} We acknowledge useful discussions with Anne Green, Edward Hardy, Hitoshi Murayama, Simon Rozier and Stephen Warren.  MF is funded by the European Research Council under the European Union's Horizon 2020 program (ERC Grant Agreement no.648680 DARKHORIZONS).  JQ and MF are supported by the UK STFC Grant ST/L000326/1 while DJEM is supported by a Royal Astronomical Society Postdoctoral Fellowship.  DJEM acknowledges hospitality of the Yukawa Institute and the UTQuest workshop, where part of this work was completed. 

\bibliographystyle{h-physrev3.bst}

\bibliography{axion_review}

\begin{thebibliography}{10}

\bibitem{FMQR}
M.~{Fairbairn}, D.~J.~E. {Marsh}, J.~{Quevillon}, and S.~{Rozier},
\newblock {In Preparation} .

\bibitem{weinberg1978}
S.~Weinberg,
\newblock \prl {\bf 40}, 223 (1978).

\bibitem{wilczek1978}
F.~Wilczek,
\newblock \prl {\bf 40}, 279 (1978).

\bibitem{1979PhRvL..43..103K}
J.~E. {Kim},
\newblock \prl {\bf 43}, 103 (1979).

\bibitem{1980NuPhB.166..493S}
M.~A. {Shifman}, A.~I. {Vainshtein}, and V.~I. {Zakharov},
\newblock Nuclear Physics B {\bf 166}, 493 (1980).

\bibitem{Zhitnitsky:1980tq}
A.~Zhitnitsky,
\newblock Sov.J . Nucl. Phys. {\bf 31}, 260 (1980).

\bibitem{1981PhLB..104..199D}
M.~{Dine}, W.~{Fischler}, and M.~{Srednicki},
\newblock Phys. Lett. B {\bf 104}, 199 (1981).

\bibitem{pecceiquinn1977}
R.~Peccei and H.~R. Quinn,
\newblock \prl {\bf 38}, 1440 (1977).

\bibitem{2008LNP...741...19S}
P.~{Sikivie},
\newblock {Axion Cosmology},
\newblock in {\em Axions}, edited by M.~{Kuster}, G.~{Raffelt}, and
  B.~{Beltr{\'a}n}, , Lecture Notes in Physics, Berlin Springer Verlag Vol.
  741, p.~19, 2008, astro-ph/0610440.

\bibitem{Marsh:2015xka}
D.~J.~E. {Marsh},
\newblock \physrep {\bf 643}, 1 (2016), 1510.07633.

\bibitem{Hogan:1988mp}
C.~J. {Hogan} and M.~J. {Rees},
\newblock Phys. Lett. B {\bf 205}, 228 (1988).

\bibitem{Sikivie:1982qv}
P.~Sikivie,
\newblock \prl {\bf 48}, 1156 (1982).

\bibitem{1987PhLB..195..361H}
D.~{Harari} and P.~{Sikivie},
\newblock Phys. Lett. B {\bf 195}, 361 (1987).

\bibitem{2016PhRvL.116c1302B}
{BICEP2 Collaboration} {\em et~al.},
\newblock \prl {\bf 116}, 031302 (2016), 1510.09217.

\bibitem{1976JPhA....9.1387K}
T.~W.~B. {Kibble},
\newblock Journal of Physics A Mathematical General {\bf 9}, 1387 (1976).

\bibitem{2015PhRvD..91f5014K}
M.~{Kawasaki}, K.~{Saikawa}, and T.~{Sekiguchi},
\newblock \prd {\bf 91}, 065014 (2015), 1412.0789.

\bibitem{1993PhRvL..71.3051K}
E.~W. {Kolb} and I.~I. {Tkachev},
\newblock \prl {\bf 71}, 3051 (1993), hep-ph/9303313.

\bibitem{1994PhRvD..49.5040K}
E.~W. {Kolb} and I.~I. {Tkachev},
\newblock \prd {\bf 49}, 5040 (1994), astro-ph/9311037.

\bibitem{Kolb:1994fi}
E.~W. {Kolb} and I.~I. {Tkachev},
\newblock \prd {\bf 50}, 769 (1994), astro-ph/9403011.

\bibitem{Kolb:1995bu}
E.~W. {Kolb} and I.~I. {Tkachev},
\newblock \apjl {\bf 460}, L25 (1996), astro-ph/9510043.

\bibitem{2007PhRvD..75d3511Z}
K.~M. {Zurek}, C.~J. {Hogan}, and T.~R. {Quinn},
\newblock \prd {\bf 75}, 043511 (2007), astro-ph/0607341.

\bibitem{2016arXiv160900208H}
E.~{Hardy},
\newblock ArXiv e-prints  (2016), 1609.00208.

\bibitem{2014JHEP...06..037D}
A.~G. {Dias}, A.~C.~B. {Machado}, C.~C. {Nishi}, A.~{Ringwald}, and
  P.~{Vaudrevange},
\newblock Journal of High Energy Physics {\bf 6}, 37 (2014), 1403.5760.

\bibitem{1983PhLB..120..127P}
J.~{Preskill}, M.~B. {Wise}, and F.~{Wilczek},
\newblock Phys. Lett. B {\bf 120}, 127 (1983).

\bibitem{1983PhLB..120..133A}
L.~F. {Abbott} and P.~{Sikivie},
\newblock Phys. Lett. B {\bf 120}, 133 (1983).

\bibitem{1983PhLB..120..137D}
M.~{Dine} and W.~{Fischler},
\newblock Phys. Lett. B {\bf 120}, 137 (1983).

\bibitem{planck_2015_params}
{Planck Collaboration} {\em et~al.},
\newblock \aap {\bf 594}, A13 (2016), 1502.01589.

\bibitem{Wantz:2009it}
O.~Wantz and E.~Shellard,
\newblock \prd {\bf 82}, 123508 (2010), 0910.1066.

\bibitem{2016PhLB..752..175B}
S.~{Borsanyi} {\em et~al.},
\newblock Physics Letters B {\bf 752}, 175 (2016), 1508.06917.

\bibitem{Borsanyi:2016ksw}
S.~Borsanyi {\em et~al.},
\newblock Nature {\bf 539}, 69 (2016), 1606.07494.

\bibitem{Press:1973iz}
W.~H. {Press} and P.~{Schechter},
\newblock \apj {\bf 187}, 425 (1974).

\bibitem{Schive:2014dra}
H.-Y. {Schive}, T.~{Chiueh}, and T.~{Broadhurst},
\newblock Nature Physics {\bf 10}, 496 (2014), 1406.6586.

\bibitem{2016arXiv161105892C}
P.~S. {Corasaniti}, S.~{Agarwal}, D.~J.~E. {Marsh}, and S.~{Das},
\newblock ArXiv e-prints  (2016), 1611.05892.

\bibitem{khlopov_scalar}
M.~Khlopov, B.~Malomed, and I.~Zeldovich,
\newblock \mnras {\bf 215}, 575 (1985).

\bibitem{Marsh:2013ywa}
D.~J.~E. {Marsh} and J.~{Silk},
\newblock \mnras {\bf 437}, 2652 (2014), 1307.1705.

\bibitem{2016JCAP...01..035T}
P.~{Tinyakov}, I.~{Tkachev}, and K.~{Zioutas},
\newblock \jcap {\bf 1}, 035 (2016), 1512.02884.

\bibitem{1986ApJ...304....1P}
B.~{Paczynski},
\newblock \apj {\bf 304}, 1 (1986).

\bibitem{Griest:1990vu}
K.~{Griest},
\newblock \apj {\bf 366}, 412 (1991).

\bibitem{Green:2016xgy}
A.~M. {Green},
\newblock \prd {\bf 94}, 063530 (2016), 1609.01143.

\bibitem{2017arXiv170102151N}
H.~{Niikura} {\em et~al.},
\newblock ArXiv e-prints  (2017), 1701.02151.

\bibitem{Tisserand:2006zx}
P.~{Tisserand} {\em et~al.},
\newblock \aap {\bf 469}, 387 (2007), astro-ph/0607207.

\bibitem{Navarroetal1997}
J.~F. Navarro, C.~S. Frenk, and S.~D. White,
\newblock \apj {\bf 490}, 493 (1997), astro-ph/9611107.

\bibitem{2017arXiv170103118O}
C.~A.~J. {O'Hare} and A.~M. {Green},
\newblock ArXiv e-prints  (2017), 1701.03118.

\bibitem{1996ApJ...461...84A}
C.~{Alcock} {\em et~al.},
\newblock \apj {\bf 461}, 84 (1996), astro-ph/9506113.

\bibitem{2002ApJ...573..597K}
A.~{Klypin}, H.~{Zhao}, and R.~S. {Somerville},
\newblock \apj {\bf 573}, 597 (2002), astro-ph/0110390.

\bibitem{2016arXiv161001639B}
G.~{Ballesteros}, J.~{Redondo}, A.~{Ringwald}, and C.~{Tamarit},
\newblock ArXiv e-prints  (2016), 1610.01639.

\bibitem{2015ARNPS..65..485G}
P.~W. {Graham}, I.~G. {Irastorza}, S.~K. {Lamoreaux}, A.~{Lindner}, and K.~A.
  {van Bibber},
\newblock Annual Review of Nuclear and Particle Science {\bf 65}, 485 (2015),
  1602.00039.

\bibitem{2016arXiv161105865T}
{The MADMAX Working Group} {\em et~al.},
\newblock ArXiv e-prints  (2016), 1611.05865.

\bibitem{2014PhRvL.113p1801A}
A.~{Arvanitaki} and A.~A. {Geraci},
\newblock \prl {\bf 113}, 161801 (2014), 1403.1290.

\end{thebibliography}

\end{document}